\documentclass[aps,showpacs,twocolumn]{revtex4}
\usepackage[dvips]{graphics,graphicx}
\begin{document}
\title{New Bloch period for interacting cold atoms in 1D optical lattices}
\author{Andrey R. Kolovsky}
\affiliation{Max-Planck-Institut f\"ur Physik Komplexer Systeme,
                             D-01187 Dresden, Germany}
\affiliation{Kirensky Institute of Physics, 660036 Krasnoyarsk, Russia}
\date{\today }

\begin{abstract}
This paper studies Bloch oscillations of ultracold atoms
in an optical lattice, in the presence of atom-atom interactions. 
A new, interaction-induced Bloch period is identified.
Analytical results are corroborated by realistic
numerical calculations.
\end{abstract}

\maketitle

The response of a quantum system to a static field has been
a longstanding problem since the early days of quantum mechanics. 
A topic of  particular interest in this
wide field is the dynamics of a quantum particle
in a periodic potential induced by a static force 
(modelling a crystal electron in an electric field).
In this system, the effect of the field manifests
in a very unintuitive way. Indeed, as already
emphasised by Bloch \cite{Bloc28} and Zener \cite{Zene34}, 
according to the predictions
of wave mechanics, the motion of electrons in a perfect
crystal should be oscillatory rather than uniform.
This phenomenon, nowadays known as Bloch oscillations (BO), 
has recently received renewed interest which was stimulated by
experiments on cold atoms in optical lattices
\cite{Daha96,Wilk96,Raiz97,Ande98}. This system
(which mimics a solid state system -- with the electrons and the crystal
lattice substituted by the neutral atoms and the optical potential,
respectively) offers unique possibilities for the experimental
study of BO and of related phenomena. In turn, these fundamentally 
new experiments have stimulated considerable progress in theory 
(see review \cite{PR}, and references therein), and it 
can be safely stated that BO in diluted quasi  
one-dimensional gases is well understood today.
Other directions of research 
focus on BO in the presence of relaxation processes
(spontaneous emission) \cite{PRA2}, BO in 2D 
optical lattices \cite{PRL3}, and BO in the presence of
atom-atom interactions (`BEC-regime') 
\cite{Berg98,Choi99,Chio00,Mors01}. The present Letter deals with
the third problem, which is approached here by an `ab initio' 
analysis of the dynamics of a system of many 
atoms. This distinguishes this work from previous studies of BO 
in the BEC regime \cite{Berg98,Choi99,Chio00}, which were based on 
the a mean field approach using a nonlinear Schr\"odinger equation.  
A new effect, so far unaddressed by these earlier studies, is predicted: 
besides the usual Bloch dynamics, the atomic oscillations
may exhibit another fundamental period, entirely 
defined  by the strength of the atom-atom interactions.

Let us first recall some results on BO in the single-particle 
case. Using the  tight-binding approximation \cite{Fuku73}, the Hamiltonian
of a single atom in an optical lattice has the form 
\begin{displaymath}
H = E_0\sum_l |l\rangle\langle l| 
-\frac{J}{2}\left(\sum_l |l+1\rangle\langle l|+h.c.\right)
\end{displaymath}
\begin{equation}
\label{1}
+dF\sum_l l |l\rangle\langle l| \;.
\end{equation}
In Eq.~(\ref{1}), $|l\rangle$ denotes the $l$th Wannier state
$\phi_l(x)$ corresponding to the  energy level $E_0$
\cite{remark1}, $J$ is the hopping matrix elements between 
neighbouring Wannier states, $d$ is the lattice period, and $F$
is the magnitude of the static force. The Hamiltonian  (\ref{1})
can be easily diagonalised, which yields the spectrum
$E_l=E_0+dFl$ (the so-called Wannier-Stark ladder) and 
the  eigenstates (Wannier-Stark states)
\begin{equation}
\label{1a}
|\psi_l\rangle=\sum_m {\cal J}_{m-l}(J/dF)|m\rangle \;,\quad
\langle x|m\rangle=\phi_m(x) \;,
\end{equation}
(here ${\cal J}_m(z)$ are the Bessel functions). 
As a direct consequence of the equidistant spectrum, 
the evolution of an arbitrary initial wave function is periodic
in time, with the Bloch period  $T_B=2\pi\hbar/dF$. In particular,
we shall be interested in the time evolution of the Bloch states
$|\psi_\kappa\rangle=\sum_l \exp(id\kappa l)|l\rangle$. 
Using the explicite expression for the Wannier-Stark states (\ref{1a}), 
it is easy to show that 
$|\psi_\kappa(t)\rangle=\exp\{-i(J/dF)\sin(d\kappa(t))\}
|\psi_{\kappa(t)}\rangle$, where $\kappa(t)=\kappa+Ft/\hbar$
(from now on $E_0=0$ for simplicity).
Note that the exponential pre-factor in the last equation
contains the same parameter $J/dF$ as the argument of the
Bessel function in Eq.~(\ref{1a}). Depending on the 
value of this parameter, the regimes of weak ($dF\ll J$) or 
strong ($dF\gg J$) static fields can be distinguished. 
In this Letter, we shall restrict ourselves 
to the strong field case, which, in some sense, is
easier to treat than the weak field regime. Indeed, for $J/dF\ll1$, the 
Wannier-Stark states practically coincide with Wannier states, and 
$|\psi_\kappa(t)\rangle\approx|\psi_{\kappa(t)}\rangle$.

A remark concerning the  characteristic values of the parameters is
at place here: In the numerical simulations below, we use scaled variables,
where $\hbar=1$, $d=2\pi$, and the energy is measured in units 
of the photon recoil energy. In typical experiments with cold 
atoms in an optical lattice, the amplitude $v$ of the 
optical potential equals few recoil energies. Then, for example, 
for $v$ equal to 10 recoil energies, the value of the dimensionless
hopping matrix element is  $J=0.0384$. 
The strength  of the static field is restricted from 
below by the condition $dF> J$, and from above by the
condition that Landau-Zener tunnelling events can be neglected.
Since the probability of Landau-Zener tunnelling is
proportional to
$\exp(-\pi\delta^2/8dFJ)$ ($\delta$ is the energy gap
separating the lowest Bloch band from the remaining part  
of the spectrum) \cite{Zene34,PR}, we have $F<30$ for $v=10$. 
\begin{figure}
\center
\includegraphics[width=8cm, clip]{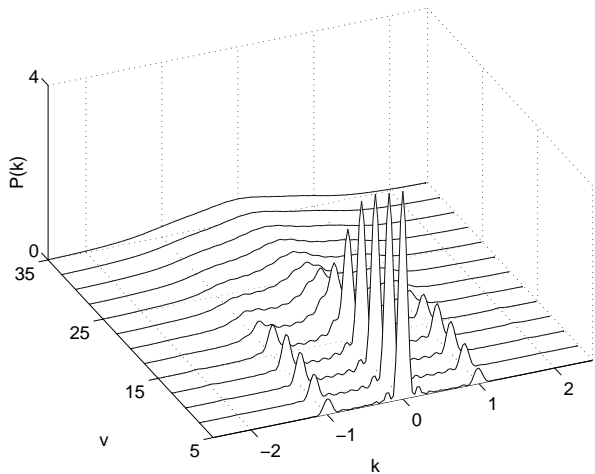}
\caption{Momentum distribution of the atoms in the 
optical lattice, for different amplitudes $v$ of the optical
potential. (The amplitude $v$ is measured in units of the recoil 
energy, the momentum $k$ in units of $2\pi\hbar/d$.) 
The figure illustrates the transition from the
SF-phase to the MI-phase as $v$ is varied ($F=0$, $L=N=7$).}
\label{fig1}
\end{figure}

We proceed with the multi-particle case. A natural
extension of the tight-binding model (\ref{1}),
which accounts for the repulsive interaction of
the atoms, is given by the Bose-Hubbard model \cite{Fish89}, 
\begin{displaymath}
H=-\frac{J}{2}\left(\sum_{l=1}^L \hat{a}^\dag_{l+1}\hat{a}_l
  +h.c.\right)
  +\frac{W}{2}\sum_{l=1}^L \hat{n}_l(\hat{n_l}-1)
\end{displaymath}
\begin{equation}
\label{2}
+2\pi F\sum_{l=1}^L l\hat{n}_l \;.
\end{equation}
In Eq.~(\ref{2}), $\hat{a}_l^\dag$ and $\hat{a}_l$ are the 
bosonic creation and annihilation operators, 
$\hat{n}_l= \hat{a}_l^\dag\hat{a}_l$ is the occupation number 
operator of the $l$th lattice site, and the parameter $W$ is
proportional to the integral over the Wannier function raised to the 
fourth power. Since the Bose-Hubbard Hamiltonian conserves the
total number of atoms $N$, the wave function of the system
can be represented in the form 
$|\Psi\rangle=\sum_{\bf n} c_{\bf n}|{\bf n}\rangle$,
where the vector ${\bf n}$, consisting of $L$ integer
numbers $n_l$ ($\sum_l n_l=N$), labels the $N$-particle
bosonic wave function constructed from $N$ Wannier functions.
(In what follows, if not stated otherwise, $|\Psi\rangle$
refers to the ground state of the system.)
As known, in the thermodynamic limit, and for $F=0$, the system 
(\ref{2}) shows a quantum phase transition from a superfluid (SF) 
to a Mott insulator (MI) phase as the ratio $J/W$ is varied 
(see \cite{Sach01} and references therein).
It is interesting to note that an indication of this transition
can already be observed in a system of few atoms \cite{Jaks98}. 
As an example, Fig.~\ref{fig1} shows the diagonal elements of the
one-particle density matrix,
\begin{equation}
\label{4}
\rho(k,k')=\langle\Psi|\hat{\Phi}^\dag(k)\hat{\Phi}(k')
 |\Psi\rangle \;,\quad 
\hat{\Phi}(k)=\sum_{l=1}^L \hat{a}_l\phi_l(k) \;,
\end{equation}
for $N=L=7$, $5\le v\le 35$, and $W=0.1\int dk \phi_l^4(k)$ 
(here, $\phi_l(k)$ are the Wannier states in the momentum
representation, and $k=p/(2\pi\hbar/d)$ is the dimensionless
momentum). Physically, this quantity corresponds
to the momentum distribution $P(k)=\rho(k,k)$ of the atoms, directly
measured in the experiment. It is seen in Fig.~\ref{fig1}
that, around $v=15$, there is a qualitative change in
the momentum distribution, in close analogy with that observed
in the experiment \cite{Grei02}. It should be noted, however,
that this qualitative change of the momentum distribution
alone does not yet prove the occurrence of a phase transition. 
A more reliable indication of a SF-MI phase transition are the  
fluctuations of the number of atoms in a single well, which drops 
from $\langle\Delta n^2\rangle\approx0.72$ at $v=5$ to
$\langle\Delta n^2\rangle\approx0$ at $v=35$ \cite{remark3}.
\begin{figure}
\center
\includegraphics[width=8cm, clip]{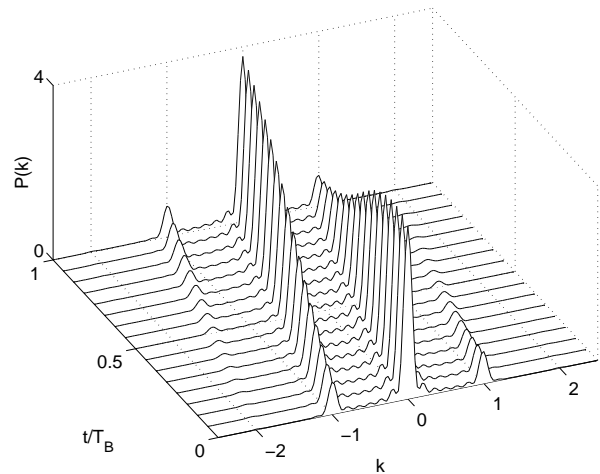}
\caption{Bloch oscillations of the atoms, induced
by the static force $F=1/2\pi$ ($v=10$). 
One Bloch period is shown.}
\label{fig2}
\end{figure}
\begin{figure}
\center
\includegraphics[width=8cm, clip]{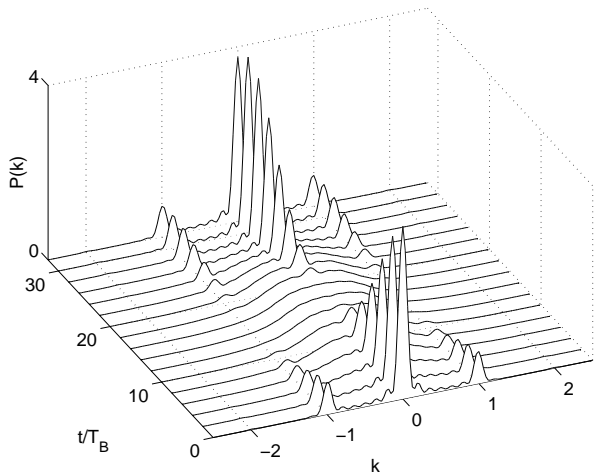}
\caption{Dephasing of Bloch oscillations due to
the atom-atom interaction. The period
$T_W=2\pi F/W$ is clearly seen. ($F=1/2\pi$, $v=10$,
$W=0.1\int dx \phi_l^4(x)=0.0324$.)}
\label{fig3}
\end{figure}

\begin{figure}
\center
\includegraphics[width=8cm, clip]{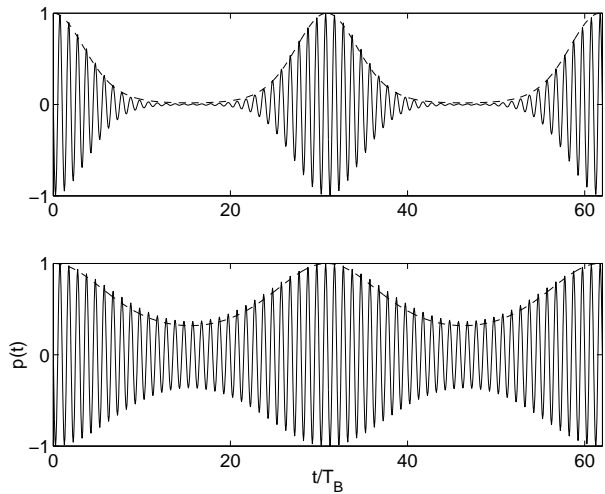}
\caption{Normalised mean momentum ($p/NJ\rightarrow p$)
as a function of time, for two different values of the
occupation number $\bar{n}=N/L$: $N=L=7$ (upper panel), and 
$N=4$, $L=14$ (lower panel). The dashed lines show the analytical
result for the envelope function in the thermodynamic limit.}
\label{fig4}
\end{figure}

Let us now discuss the effect of the static force.
Figure \ref{fig2} shows the dynamics of the momentum 
distribution of the atoms (which were initially in SF-phase)
in presence of a force $F=1/2\pi$ \cite{remark2}. This numerical
simulation illustrates atomic BO as observed in laboratory 
experiments \cite{Daha96, Mors01}.
It is seen that  after one Bloch period the initial
momentum distribution practically coincides with the
final distribution. A small difference, which can be 
noticed  by closer inspection of the figure, is
obviously due to the atom-atom interaction \cite{remark4}. 
This difference becomes evident once the system evolved 
over several Bloch periods. In  Fig.~\ref{fig3}, 
the momentum distribution $P(k)$ at integer multiples of the
Bloch period is shown. A periodic change of the distribution
from SF to MI-like and back is clearly seen.
(The use of the term `MI-like' stresses the fact that the 
variance $\langle\Delta n^2\rangle$ does not change as
time evolves.) 
In addition to Fig.~\ref{fig2} and Fig.~\ref{fig3}, 
Fig.~\ref{fig4} depicts the mean momentum $p(t)$ of the atoms
for two different values of the occupation number
$\bar{n}=N/L$ (number of atoms per lattice cite) -- 
$\bar{n}=1$ (upper panel) and $\bar{n}=2/7$ (lower panel). 
As to be expected, the dynamics of the system depends on the 
value of $\bar{n}$, and for a larger occupation number the 
deviations of many-particle BO from 
the non-interacting result $p(t)=NJ\sin(2\pi Ft)$ 
becomes larger.

Our explanation for the numerical results is the following.
It is convenient to treat the atom-atom interaction as
a perturbation. Let us denote by $U_0(t)$ the evolution
operator of the system for $W=0$, by  $U(t)$ the evolution
operator for $W\ne0$, and by $U_W(t)$ the evolution operator defined
by the decomposition $U(t)=U_W(t)U_0(t)$. 
Since $U_0(T_B)$ is the identity matrix, one has
to find  $U_W(T_B)$ to reproduce the result of
Fig.~\ref{fig3}. In the interaction representation, 
the formal solution for $U_W(T_B)$ has the form
\begin{equation}
\label{5}
U_W(T_B)=\widehat{\exp}\left(-i\frac{W}{2}
\int_0^{T_B} dt U_0^\dag(t)
\sum_{l=1}^L \hat{n}_l(\hat{n_l}-1)U_0(t)\right) \;,
\end{equation}
where the hat over the exponential denotes time ordering.
Now we make use of the above strong-field condition
$dF>J$. Under this premise the Wannier states are the eigenstates 
of the atom in the optical lattice [see Eq.~(\ref{1a})]. 
In the multi-particle case this means that 
the Fock states $|{\bf n}\rangle$ are the eigenstates of
the system (\ref{2}) and, thus, that the operator $U_0(t)$
is diagonal in the Fock state basis. Then the integral
in Eq.~(\ref{5}) can be calculated explicitely, which yields 
\begin{equation}
\label{6}
\langle {\bf n}|U_W(T_B)|{\bf n}\rangle=
\exp\left(-i\frac{W}{2F}\sum_{l=1}^L\langle {\bf n}|
\hat{n}_l(\hat{n_l}-1)|{\bf n}\rangle \right) \;.
\end{equation}
Finally, by noting that the quantity 
$\langle {\bf n}|\hat{n}_l(\hat{n_l}-1)|{\bf n}\rangle$
is always an even integer, one comes to the conclusion
that, besides the Bloch period, there is additional period,
\begin{equation}
\label{7}
T_W=2\pi F/W \;,
\end{equation}
which characterises the dynamics of the system.

Further analytical results can be obtained 
if we approximate the ground state of the system for $F=0$ by the
product of $N$ Bloch waves with quasimomentum $\kappa=0$, i.e.,
\begin{equation}
\label{8}
|\Psi\rangle=\frac{1}{\sqrt{N!}}\left(
\frac{1}{\sqrt{L}}\sum_{l=1}^L \hat{a}^\dag_l\right)^N  
|0\ldots0\rangle \;.
\end{equation}
Indeed, let us consider, for example, the dynamics of the mean momentum. 
Using the interaction representation (now with respect to the
Stark energy term) the mean momentum is given by
\begin{equation}
\label{9}
p(t)=J\; {\rm Im}\left(\langle \Psi U^\dag_W(t)|
\sum_{l=1}^L \hat{a}^\dag_{l+1}\hat{a}_l
|U_W(t) \Psi \rangle e^{-i2\pi Ft}\right) \;,
\end{equation}
where $U_W(t)$ is the continuous-time version ($T_B\rightarrow t$) 
of the diagonal unitary matrix (\ref{6}).
Substituting  Eq.~(\ref{8}) and Eq.~(\ref{6}) into Eq.~(\ref{9}),
we obtain the following exact expression,
\begin{equation}
\label{10}
\frac{p(t)}{NJ}=\frac{L}{N} {\rm Im}\left(
\sum_{n,n'} n {\cal P}(n,n')e^{i(n'-n+1)Wt}e^{-i2\pi Ft}\right) \;,
\end{equation}
where ${\cal P}(n,n')$ is the joint probability to find $n$ and $n'$
atoms in two neighbouring wells. In the thermodynamic
limit $N,L\rightarrow\infty$, $N/L=\bar{n}$, the function
${\cal P}(n,n')$ factorises into a product of the Poisson distributions
${\cal P}(n)=\bar{n}^n\exp(-\bar{n})/n!$, and the double sum in 
Eq.~(\ref{10}) converges to the positive periodic function,
$f(t)=\exp(-2\bar{n}[1-\cos(Wt)])$,
indicated in Fig.~\ref{fig4} by the dashed line. Good
agreement between the envelope of $p(t)$ and the dashed line proves
that in the numerical simulation presented above the convergence
was indeed achieved.


In summary, Bloch oscillations of interacting cold atoms 
have been studied, both numerically and analytically. 
We have shown that in the strong field regime 
atom-atom interactions cause the reversible
dephasing of Bloch oscillations. As a result, the momentum 
distribution of the atoms changes periodically from SF to MI-like 
distributions, with a period given by Eq.~(\ref{7}).
Using the original (unscaled) parameters, this period reads 
$T_W=(dF/W)T_B=2\pi\hbar/W$, where $W$ is the 
strength of the atom-atom interactions. Since the momentum
distribution can be measured easily in the laboratory 
experiment, this effect suggests an alternative method 
for studying atom-atom interactions by 
applying a static force to the system. 
It is worth to stress one more time that the reported 
result is valid only in the strong
field limit, $dF> J$. If this condition is violated, the
evolution operator (\ref{6}) is no more a diagonal matrix,
and the system dynamics get significantly more complicated.
An analysis of this latter case will be presented elsewhere.

Discussions with A.~Buchleitner are gratefully acknowledged.



\begin{thebibliography}{10}
\bibitem{remark1}
The Wannier states (which should not be confused with
Wannier-Stark states) are defined as the Fourier coefficients 
of the Bloch states
over the quasimomentum. For the considered range
the $v$, the $l$th Wannier state is essentially
a Gaussian centred at the $l$th potential well of the
optical lattice.
\bibitem{remark3}
A SF-MI transition can also be tracked by calculating
the entropy $S=-\sum_{\bf n} |c_{\bf n}|^2 \log |c_{\bf n}|^2$,
which changes from $S\sim\log {\cal N}$ (${\cal N}$ is the
dimension of the Hilbert space) in the SF-phase, to $S\approx0$ in the
MI-phase.
\bibitem{remark2}
In all numerical simulations periodic boundary conditions 
were used, i.e., the site $l=L+1$ is identified with $l=1$.
Note that in the case $F\ne0$ this can be done
only after elimination of the static term, what is
achieved in the interaction representation.
Thus, the dynamics of the system (\ref{2}) was actually calculated
on the basis of the time-dependent Hamiltonian
$\widetilde{H}(t)=-(J/2)\left(\exp(-i2\pi Ft)\sum \hat{a}^\dag_{l+1}
\hat{a}_l +h.c.\right)+(W/2)\sum \hat{n}_l(\hat{n_l}-1)$.
\bibitem{remark4}
One gets exact coincidence between the initial and final
distributions only for $W=0$. It might also be useful to mention
two obvious facts concerning this case. First, for $W=0$ the
momentum distribution $P(k)$ calculated for the 
many atoms system coincides with that obtained within the
single-particle approach. Second, within the framework of our 
present tight-binding model, the function $P(k,t/T_B)$
does not depend on the particular value of the static
force $F$.
\end{thebibliography}
\end{document}